\documentclass[useAMS,usenatbib]{mn2e}

\usepackage{graphicx}
\usepackage{times}
\usepackage{natbib}
 
\def\lesssim{\mathrel{\hbox{\rlap{\hbox{\lower4pt\hbox{$\sim$}}}\hbox{$<$}}}}
\def\gtrsim{\mathrel{\hbox{\rlap{\hbox{\lower4pt\hbox{$\sim$}}}\hbox{$>$}}}}

\def\del#1{{}}
\def\pacman{{\it Pacman }}
\def\be{\begin{equation}}
\def\ee{\end{equation}}
\def\RM{{\rm RM}}

\title[Pacman II: Application]{\textbfit{Pacman} (II):
Application and Statistical Characterisation of Improved RM Maps} 
\author[C. Vogt, K. Dolag, T. A. En\ss lin]
{C. Vogt$^{1}$\thanks{E-mail: cvogt@mpa-garching.mpg.de (CV);
kdolag@pd.astro.it (KD); ensslin@mpa-garching.mpg.de (TAE)},
K. Dolag$^{2}$ and T. A. En\ss lin$^{1}$ \\ $^{1}$Max-Planck-Institut
f\"{u}r Astrophysik, Karl-Schwarzschild-Str.1, Postfach 1317, 85741
Garching, Germany \\ $^{2}$Dipartimento di Astronomia, Universita di
Padova, vicolo dell'Osservatorio 2, 35122 Padova, Italy}
 
\begin{document}

\date{Accepted ???. Received ???; in original form ???}

\pagerange{\pageref{firstpage}--\pageref{lastpage}} \pubyear{0000}

\maketitle

\label{firstpage}

\begin{abstract} 
We proposed a new method -- \textit{Pacman} -- to calculate
Faraday rotation measure (RM) maps from multi-frequency polarisation
angle data (Dolag et al.) in order to avoid the so-called
$n\upi$-ambiguity. Here, we apply our \pacman algorithm to two
polarisation data sets of extended radio sources in the Abell 2255 and
the Hydra~A cluster, and compare the RM maps obtained using \pacman to
RM maps obtained employing already existing methods. Thereby,
we provide a new high quality RM map of the Hydra north lobe which is
in a good agreement with the existing one but find significant
differences in the case of the south lobe of Hydra A. We demonstrate
the reliability and the robustness of \emph{Pacman}. In order to study
the influence of map making artefacts, which are imprinted by wrong
solutions to the $n\upi$-ambiguities, and of the error treatment of
the data, we calculated and compared magnetic field power spectra from
various RM maps. The power spectra were derived using the method
recently proposed by \citet{2003A&A...401..835E}. We demonstrate the
sensitivity of statistical analysis to artefacts and noise in the RM
maps and thus, we demonstrate the importance of an unambiguous
determination of RM maps and an understanding of the nature of the
noise in the data. We introduce and perform statistical tests to
estimate the quality of the derived RM maps, which demonstrate the
quality improvements due to \pacman.
\end{abstract}


\section{Introduction} \label{sec:intro}
Magnetic fields are ubiquitous throughout the universe. One method to
study them is the analysis of the Faraday rotation effect. This effect
describes that if polarised radiation traverses a magnetised plasma
the plane of polarisation of the radiation is rotated. The change in
polarisation angle is proportional to the squared wavelength
$\lambda^2$. The proportionality constant is called rotation measure
(RM). Various methods have been proposed in order to calculate RM maps
from multi-frequency polarisation data sets
\citep[e.g.][]{1975A&A....43..233V, 1975MNRAS.173..553H,
1979A&A....78....1R, 2001MNRAS.328..623S}. The difficulty involved in
this calculation arises from the fact that observations constrain the
polarisation angle $\varphi$ only up to additions of $\pm n\upi$. This
leads to the so-called $n\upi$-ambiguity.

In a previous paper \citep[][ hereafter Paper I]{paper1}, we proposed
a new approach for the unambiguous determination of RM and of the
intrinsic polarisation angle $\varphi^0$ at the observed, polarised
radio source from multi-frequency polarisation data sets. The proposed
algorithm uses a global scheme in order to solve the
$n\upi$-ambiguity. It assumes that if regions exhibit small
polarisation angle gradients between neighbouring pixels in all
observed frequencies simultaneously then these pixels can be
considered as connected. Information about one pixel can be used for
neighbouring ones and especially the solution to the $n\upi$-ambiguity
should be the same.

Faraday rotation measure maps are analysed in order to get insight
into the properties of the RM producing magnetic fields such as field
strengths and correlation lengths. Artefacts in RM maps which result
from $n\upi$-ambiguities can lead to misinterpretation of the
data. Recently, \citet{2003A&A...401..835E} proposed a method to
calculate magnetic power spectra from RM maps and to estimate magnetic
field properties. They successfully applied their method to RM maps
\citep{2003A&A...412..373V} and realised that map making artefacts and small
scale pixel noise have a tremendous influence on the shape of the
magnetic power spectra on large Fourier scales -- small real space
scales. Thus, for the application of these kind of statistical methods
it is desirable to produce unambiguous RM maps with as little noise
as possible. This and similar applications are our motivations for
designing \pacman (Paper I) and to quantify its performance (this
Paper).

In order to detect and to estimate the correlated noise level in RM
and $\varphi^0$ maps, \citet{2003ApJ...597..870E} recently proposed a
gradient vector product statistic. It compares RM gradients and
intrinsic polarisation angle gradients and aims to detect correlated
fluctuations on small scales, since RM and $\varphi^0$ are both
derived from the same set of polarisation angle maps.

In Paper I, we describe in detail the idea and the implementation of
the \textit{Pacman}\footnote{The computer code for \textit{Pacman}
will be publicly available at {\tt
http://dipastro.pd.astro.it/\~{}cosmo/Pacman}}
algorithm. Furthermore, we presented a test of the algorithm
on artificially generated maps where we demonstrated that \pacman
yields the right solution to the $n\pi$-ambiguity. Here, we want to
apply this algorithm to polarisation observation data sets of two
extended polarised radio sources located in the Abell 2255
\citep{2002taiwan.conf} cluster and the Hydra cluster
\citep{1993ApJ...416..554T}. This is described in
Sect.~\ref{sec:application}. Using these polarisation data, we
demonstrate the stability of our \pacman algorithm and compare RM maps
obtained from a standard fit algorithm. A ``standard fit'' algorithm
is considered to be the algorithm which performs a pixel by pixel
least squares fit using the methods suggested by
\citet{1975A&A....43..233V, 1975MNRAS.173..553H, 1979A&A....78....1R}.

We demonstrate the importance of the unambiguous determination of RMs
for a statistical analysis by applying the statistical approach
developed by \citet{2003A&A...401..835E} to the RM maps in order to
derive the power spectra and strength of the magnetic fields in the
intra-cluster medium. We also discuss the influence of error treatment
in the analysis. The philosophy of this statistical analysis and the
calculation of the power spectra is briefly outlined in
Sect.~\ref{sec:powerspectra} whereas the results of the application to
the RM maps are presented in Sect.~\ref{sec:application}. 

In addition, we apply the \textit{gradient vector product statistic V}
as proposed by \citet{2003ApJ...597..870E} in order to detect map
making artefacts and correlated noise. The concept of this statistic
is briefly explained in Sect.~\ref{sec:avstat} and in
Sect.~\ref{sec:application}, it is applied to the data. After the
discussion of our results, we give our conclusions and lessons learned
from the data during the course of this work in
Sect.~\ref{sec:conclusion}.

Throughout the rest of the paper, we assume a Hubble constant of
H$_{0} = 70$ km s$^{-1}$ Mpc$^{-1}$, $\Omega_{m} = 0.3$ and
$\Omega_{\Lambda} = 0.7$ in a flat universe. The notation used follows
Paper I.

\section{Statistical Analysis of RM maps} 

\subsection{Gradient Vector Product Statistic} \label{sec:avstat}
\citet{2003ApJ...597..870E} introduced a \textit{gradient vector
product statistic V} to reveal correlated noise in the data. Observed
RM and $\varphi^0$ maps will always have some correlated fluctuations
on small scales, since they are both calculated from the same set of
polarisation angle maps, leading to correlated fluctuation in both
maps. The noise correlation between $\varphi^0$ and RM errors is an
anti-correlation of linear shape and can be detected by comparing the
gradients of these quantities. A suitable quantity to detect
correlated noise between $\varphi^0$ and RM is therefore
\begin{equation}\label{eq:Vstat}
V=\frac{\int d^2x \,\bmath{\nabla}\RM(\bmath{x}) \cdot
\bmath{\nabla}\varphi^0(\bmath{x})} {\int d^2x\,
|\bmath{\nabla}\RM(\bmath{x})| |\bmath{\nabla}\varphi^0(\bmath{x})|},
\end{equation}
where $\bmath{\nabla}\RM(\bmath{x})$ and
$\bmath{\nabla}\varphi^0(\bmath{x})$ are the gradients of RM and
$\varphi^0$. A map pair, which was constructed from a set of
independent random polarisation angle maps $\varphi(k)$, will give a
$V\ga-1$. A map pair without any correlated noise will give a
$V\approx0$. Hence, the statistic \textit{V} is suitable to detect
especially correlated small scale pixel noise.

The denominator in Eq.~(\ref{eq:Vstat}) is the normalisation and
enables the comparison of \textit{V} for RM maps of different sources,
since $V$ is proportional to the fraction of gradients which are
artefacts. Since, we especially want to compare the quality between
maps of the same source calculated using the two different algorithms,
it is useful to introduce the unnormalised quantity
\begin{equation}
\tilde{V}=\int d^2x\,\bmath{\nabla}\RM(\bmath{x}) \cdot
\bmath{\nabla}\varphi^0(\bmath{x}). 
\end{equation}
The quantity $\tilde{V}$ gives the absolute measure of correlation
between the gradient alignments of RM and $\varphi^0$. The smaller
this value is the smaller is the total level of correlated noise in the
maps.

\subsection{Error Under or Over Estimation}\label{sec:s_statistic}
One problem, we were faced with during the course of our statistical
analysis is the possibility that the measurement errors are under or
overestimated. Such a hypothesis can be tested by performing a
reduced $\chi^2_{\nu}$ test which is considered to be a measure for
the goodness of each least squares fit and calculates as follows
\begin{equation}
\chi^2_{\nu_{ij}} = \frac{1}{\nu} \sum_{k = 1} ^f \frac{\left[
\varphi_{{\rm {obs}}_{ij}} (k) - (\RM_{ij}\,\lambda_k ^2 +
\varphi_{ij} ^0) \right]^2}{\sigma_{k_{ij}}^2} = \frac{s_{ij}^2}
{\langle \sigma_{k_{ij}} \rangle_k},
\end{equation}
where $\nu = f - n_c$ is the number of degrees of freedom and
$f$ is the number of frequencies used and $n_c$ is the number of model
parameters in our case, $n_c = 2$. The parameter $s^2$ is the variance
of the fit and $\langle \sigma_{k_{ij}} \rangle_k$ is the weighted
average of the individual variances:
\begin{equation}
\langle \sigma_{k_{ij}} \rangle_k = \left[ \frac{1}{f} \sum_{k =
1}^ f \frac{1}{\sigma_{k_{ij}}^2}  \right]^{-1}
\end{equation}

If one believes in the assumption of Gaussian noise, that the data are
not corrupted and that the linear fit is an appropriate model then any
statistical deviation from unity of the map average $\overline{\chi}^2
_{\nu}$ of this value indicates an under or over error estimation. For
a $\overline{\chi}_{\nu}^2 > 1$, the errors have been underestimated
and for a $\overline{\chi}_{\nu}^2 < 1$, they have been overestimated.

Unfortunately this analysis does not reveal in its simple form at
which frequency the errors might be under or overestimated. However,
one can test for the influence of single frequencies by leaving out
the appropriate frequency for the calculation of the RM map and
comparing the resulting $\chi^2_{\nu_{ij}}$ values with the original
one.

It is advisable to evaluate the $\chi^2_{\nu_{ij}}$-maps in order to
locate regions of too high or too low values for $\chi^2_\nu$ by eye.

\subsection{Magnetic Power Spectra} \label{sec:powerspectra}
Our \pacman algorithm is especially useful for the determination of RM
maps of extended radio sources. RM maps of extended radio sources are
frequently analysed in terms of correlation length and the RM
producing magnetic field strength. One analysis relying on statistical
methods in order to derive magnetic field power spectra was recently
developed by \citet{2003A&A...401..835E}. They successfully applied
their method to existing RM maps of radio sources located in Abell
2634, Abell 400 and Hydra~A \citep{2003A&A...412..373V}. In the course
of their work, they realised that their method is sensitive to map
making artefacts and pixel noise, which can lead to misinterpretation
of the data at hand. Therefore this method is a good opportunity to
study the influences of noise and artifacts on the resulting power
spectra.

The observational nature of the data is taken into account by
introducing a window function, which can be interpreted as the
sampling volume. Especially, a noise reducing data weighting scheme
can be introduced to account for observational noise. One reasonable
choice of weighting is to introduce a factor $\sigma_0
^{\RM}/\sigma^{\RM}_{ij}$, which we call simple window
weighting. Another possible weighting scheme is a thresholding scheme
described by $1/(1+ \sigma_0 ^{\RM}/\sigma^{\RM}_{ij})$, which means
that the noise below a certain threshold $\sigma_0 ^{\RM}$ is
acceptable and areas of higher noise are down-weighted.

Note, that magnetic power spectra, which are calculated using the
approach explained above, are used as valuable estimator for magnetic
field strengths and correlation lengths. However, since the spectra
are shaped by and are very sensitive to small scale noise and map
making artefacts, we use them here to quantify the influence of noise
and artefacts rather than as an estimator for characteristic
properties of the RM producing magnetic fields.

\section{Application to data} \label{sec:application}
For the calculation of what we call further on the standard fit RM
maps, we used the $\chi^2$ method suggested by
\citet{1975A&A....43..233V} and \citet{1975MNRAS.173..553H}. Also for
the individual RM fits, we adopted this method. We used as an upper
limit for the RM values $|\RM|_{ij} = \RM^{\max} \pm
3\sigma^{\RM}_{ij}$, where $\sigma^{\RM}_{ij}$ was calculated
following Eq.(5) in Paper I. The least-squares fit for the individual
data points were always performed as error weighted fits as described
by Eq.~(3) and (4) in Paper I, if not stated otherwise. For the
\pacman calculations, we used for the free parameters $\alpha = \beta
= 1$ in Eq.~(7) in Paper I. 

However, we also used the method suggested
by \citet{1979A&A....78....1R} as standard fit algorithm. The results
did not change substantially.

\begin{figure*}
\includegraphics[width=0.9\textwidth]{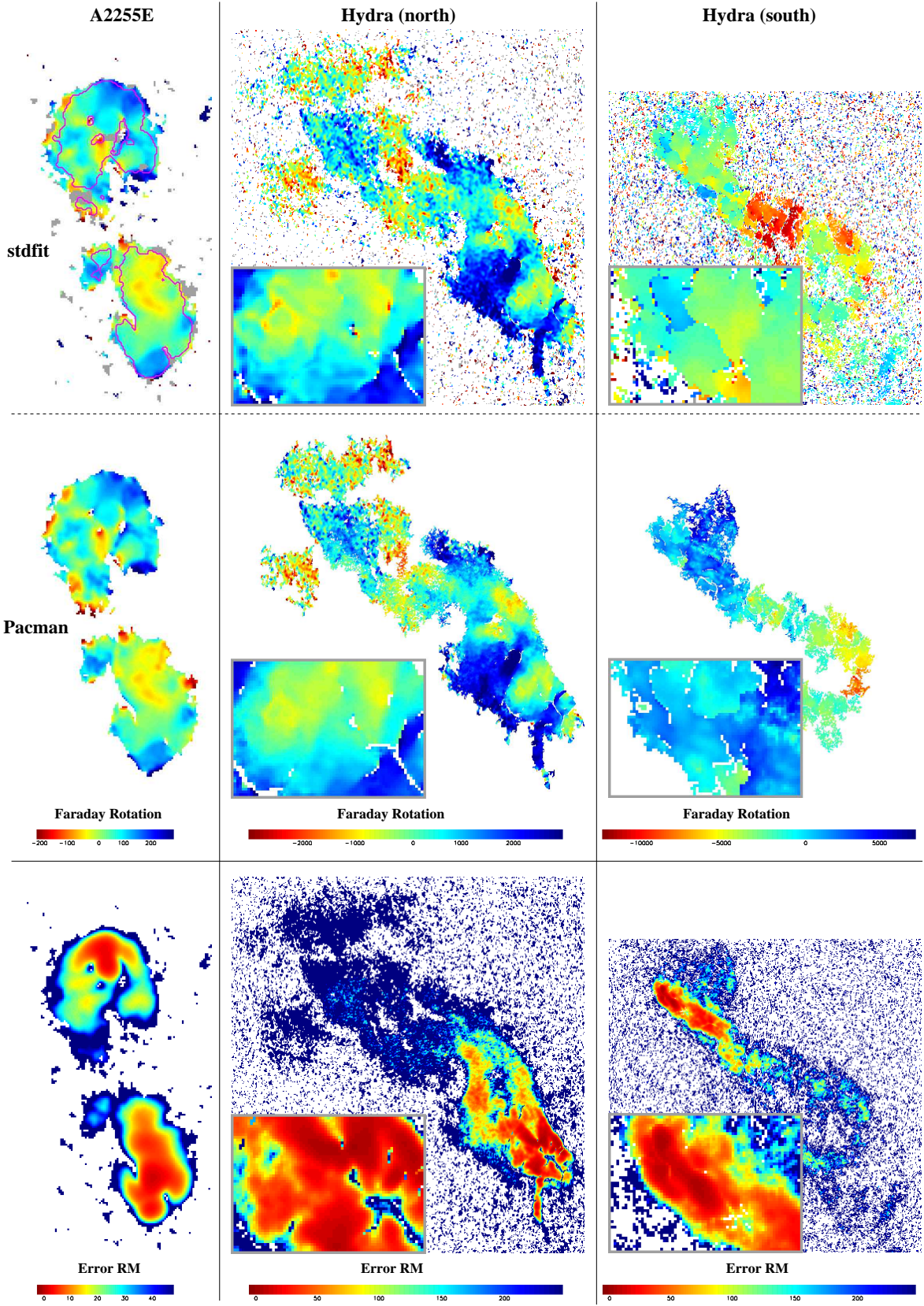}
\caption{Comparison of the standard fit RM maps (upper panels) and the
\pacman RM maps (middle panels) for A2255E (left column), the north
(middle column) and the south lobe (right column) of Hydra~A. RM
values not represented by the colour bar are coloured in grey. Panels
at the bottom exhibit the respective $\sigma^{\RM}$-maps. For the
standard fit RM map of A2255E a $\sigma_k ^{\max} = 25\degr$ and
$k_{min}= f = 4$ was used, contours indicate the area covered
by a standard fits RM map with $\sigma_k ^{\max} = 10\degr$. The
parameters used for the \pacman RM map of A2255E are $\sigma_k ^{\max}
= 25\degr$, $\sigma^{\Delta}_{\max} = 25\degr$, $g$ = 1.2 and
$k_{min}= f = 4$; for the standard fit RM maps of Hydra~A are
$\sigma_k ^{\max} = 25\degr$ and $k_{min}= f = 4$ (at 8 GHz); for
the \pacman RM map of Hydra (north) are $\sigma_k ^{\max} =
30\degr$, $k_{min}= 4$, $f = 5$, $\sigma^{\Delta}_{\max} =
25\degr$ and $g$ = 1.5; for the \pacman RM map of Hydra (south) are
$\sigma_k ^{\max} = 35\degr$, $k_{min} = f = 5$,
$\sigma^{\Delta}_{\max} = 35\degr$ and $g$ = 2.0.}
\label{fig:rmmaps}
\end{figure*}

\subsection{Abell 2255E}
The Abell cluster 2255, which has a redshift of 0.0806
\citep{1999ApJS..125...35S}, has been studied by
\citet{1995ApJ...446..583B} and \citet{1997A&A...317..432F}. The
polarised radio source B1713+641, which we call hereafter A2255E, has
a two sided radio lobe structure and is not directly located in the
cluster centre. Polarisation observations were performed using the
Very Large Array (VLA) at 4535, 4885, 8085, and \mbox{8465 MHz}. The
data reduction was done with standard AIPS (Astronomical Imaging
Processing Software) routines (Govoni et al., in prep.). The
polarisation angle maps and their standard error maps for the four
frequencies were kindly provided to us by Federica Govoni. A
preliminary analysis of an RM map of this source in order to determine
the properties of the magnetic field in the intra-cluster gas in Abell
2255 is presented in \citet{2002taiwan.conf}.

Using the polarisation angle data, an RM map employing the standard
fit algorithm was calculated, where the maximal allowed error in
polarisation angle was chosen to be \mbox{$\sigma_k ^{\max}=25
\degr$}. The resulting RM map is shown in the upper left panel of
Fig \ref{fig:rmmaps}. The overlaid contour indicates the area which
would be covered if polarisation angle errors were limited by $\sigma_k
^{\max}=10 \degr$. For these calculations, we assumed
\mbox{$\RM^{\max} = 1\,500$ rad m$^{-2}$}.

As can be seen from this map, typical RM values range between
\mbox{-100 rad m$^{-2}$} and \mbox{+210 rad m$^{-2}$}. Since this
cluster is not known to inhabit a cooling core in its centre
\citep{1997A&A...317..432F}, these are expected values. However, the
occurrence of RM values around 1000 rad m$^{-2}$, which can be seen as
grey areas in the standard fit map of A2255E, might indicate that for
these areas the $n\upi$-ambiguity was not properly solved. One reason
for this suspicion is found in the rapid change of RM values occurring
between 1 or 2 pixel from 100 to \mbox{1000 rad m$^{-2}$}. All these
jumps have a $\Delta\RM$ of about \mbox{1000 rad m$^{-2}$} indicating
$n\upi$-ambiguities between 4 GHz and 8 GHz which can be
theoretically calculated from $\Delta\RM = \upi/(\lambda_2 ^2 -
\lambda_1 ^2)$. Another important point to note is that parts of the
grey areas lie well within the $\sigma_k ^{\max}=10 \degr$ contour,
and hence, contribute to the results of any statistical analysis of
such a parametrised RM map.

Therefore, the polarisation data of this source provide a good
possibility to demonstrate the robustness and the reliability of our
proposed algorithm. The RM map calculated by \pacman is shown in the
middle left panel. The same numbers for the corresponding parameters
were used; $\sigma_k ^{\max}=25 \degr$, \mbox{$\RM^{\max} = 1\,500$
rad m$^{-2}$}.

An initial comparison by eye reveals that the method used in the
standard fit procedure produces spurious RMs in noisy regions
(manifested as grey areas in the upper left panel of
Fig. \ref{fig:rmmaps}) as previously mentioned while the RM map
calculated by \pacman shows no such grey areas. The apparent jumps of
about \mbox{1000 rad m$^{-2}$} observed in the standard fit map have
disappeared in the \pacman map. Note, that in principle by realising
that these steps are due to $n\pi$-ambiguities, one could re-run the
standard fit algorithm with a lower $\RM_{\max}$ and this wrong
solutions would disappear. However, by doing this a very strong bias
is introduced since there might also be large RM values which are
real. This bias can be easily relaxed by using \textit{Pacman}.

\begin{figure}
\includegraphics[width=0.5\textwidth]{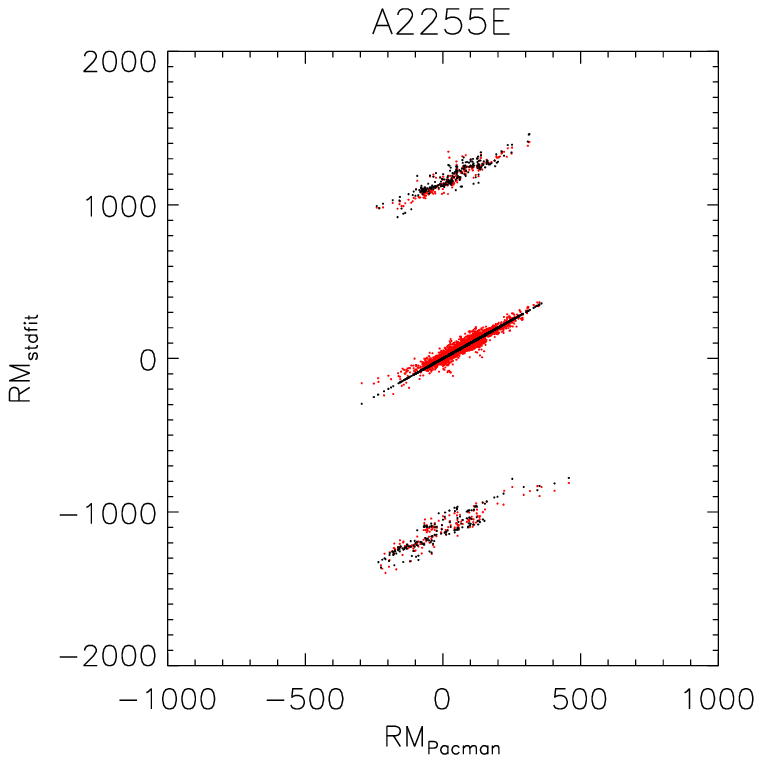}
\caption{A pixel-by-pixel scatter plot of $\RM_{\rm {stdfit}}$ values
obtained using the standard fit on the $y$-axis versus the $\RM_{\rm
{Pacman}}$ values calculated from the polarisation data of
Abell 2255 by employing our new algorithm \emph{Pacman}. Black points
represent results from an error weighted standard least-squares fit
and red points from a non-error weighted standard fit. The
points for $\RM_{\rm {stdfit}}$ values of around $\pm$ 1000
${\mathrm {rad}}\,{\mathrm m}^{-2}$ are artefacts of a wrong solution
of the $n\upi$-ambiguity.}
\label{fig:compa2255e}
\end{figure}

A pixel-by-pixel comparison is shown in Fig. \ref{fig:compa2255e}. In
this figure the $\RM_{\rm {stdfit}}$ values obtained for pixels using
a standard fit are plotted on the $y$-axis against the $\RM_{\rm
{Pacman}}$ values obtained at the same corresponding pixel locations
using the \pacman algorithm on the $x$-axis. The black points
represent the error weighted standard fit whereas the red points
represent the non-error weighted standard fit. Note that \pacman
applied only an error weighted least-squares fit to the data. From
this scatter plot, one can clearly see that both methods yield same RM
values for most of the points which is expected from the visual
comparison of both maps. However, the points for $\RM_{\rm
{stdfit}}$ values at around \mbox{$\pm$ 1000 rad m$^{-2}$} are due to
the wrong solution of the $n\upi$-ambiguity found by the standard fit
algorithm. 

For the demonstration of the effect of $n\upi$-ambiguity artefacts and
pixel noise on statistical analysis, we determined the cluster
magnetic field power spectra by employing the approach briefly
described in Sect. \ref{sec:powerspectra}. For detailed discussion and
description of the application of the method to data, we refer the
reader to \citet{2003A&A...412..373V}. In the calculation, we assumed
that the source plane is parallel to the observer plane. Please note,
that the application of an RM data filter in order to remove bad data
or data which might suffer from a wrong solution to the
$n\upi$-ambiguity as described by \citet{2003A&A...412..373V} is here
not necessary and not desirable. This is on the one hand due to the
use of an error weighting scheme in the window function to suppress
bad data and on the other hand due to the wish to study the influence
of noise and map making artefacts on the power spectra.

The RM area used for the calculation of any power spectrum is the RM
area which would be covered by the \pacman algorithm while using the
same set of parameter $\sigma_{k}^{\max}$, $\RM^{max}$. This ensures
that pixels on the border or noisy regions of the image not associated
with the source -- as seen in the standard fit RM map of A2255E in the
upper left panel of Fig.~\ref{fig:rmmaps} -- are not considered for
the analysis. (The same philosophy also applies for the discussion of
the Hydra data in Sect.~\ref{sec:hydranorth} and
Sect.~\ref{sec:hydrasouth}).

\begin{figure}
\includegraphics[width=0.5\textwidth]{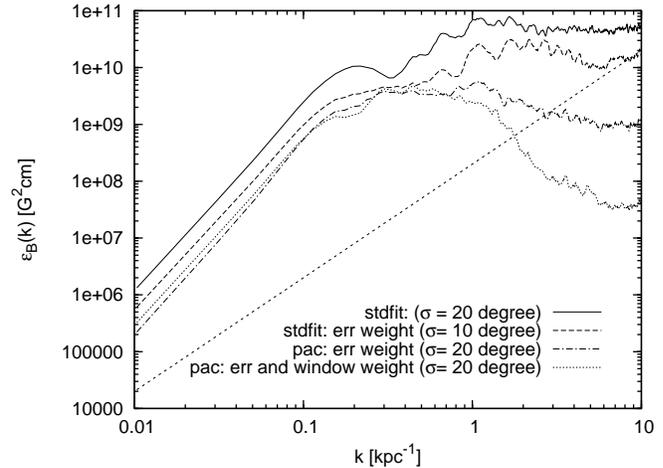}
\caption{Various power spectra for the cluster Abell 2255 determined
from different RM map making scenarios are shown. The solid line
represents a power spectrum calculated from a standard fit RM map with
$\sigma_k ^{\max}=20\degr$ where no error weighting was applied. The
dashed power spectrum was obtained from a standard fit RM map with
$\sigma_k ^{\max}=10\degr$ where error weighting was applied to the
standard fit. The dashed-dotted power spectrum was determined from a
\pacman RM map with $\sigma_k ^{\max}=20\degr$ and error weighting was
applied to the RM fits. The dotted line represents a power spectrum
calculated from a \pacman RM map as above, but additionally a simple
window weighting was applied for the determination of the power
spectra. For comparison, the power spectra of pure white noise
is plotted as a straight dashed line.}
\label{fig:psp2255e}
\end{figure}

The power spectra calculated for various map making scenarios are
shown in Fig. \ref{fig:psp2255e}. The solid and dashed line represent
power spectra calculated from standard fit RM maps. The solid line was
calculated from an RM map which was obtained by using \mbox{$\sigma_k
^{\max}=20 \degr$} where no error weighting was applied to the
standard fit. The possible window weighting as introduced in
Sect. \ref{sec:powerspectra} was also not applied. This scenario is
therefore considered as a worst case. The dashed line was calculated
from an RM map employing an error weighted standard fit while only
allowing errors in the polarisation angle of $\sigma_k ^{\max} =
10\degr$. Again no window weighting was applied. From the comparison
of these two power spectra alone, one can see that the large
$k$-scales -- and thus small real space scales -- are sensitive to
pixel-noise. Therefore, the application of an error weighted least
square fit in the RM calculation leads to a reduction of noise in the
power spectra. Note that in these two power spectra, the influence of
the $n\upi$-artefacts are still present even in the error weighted
standard fit RM map.

The two remaining power spectra in Fig.~\ref{fig:psp2255e} allow us to
investigate the influence of the $n\upi$-artefacts. The dashed-dotted
line represents the power spectra as calculated from a \pacman RM map
allowing $\sigma_k ^{\max} = 20\degr$ performing an error weighted
fit. Again, no window weighting was applied to this calculation. One
can clearly see that there is one order of magnitude difference
between the error weighted standard fit power spectra and the \pacman
one at large $k$-values. Since we also allow higher noise levels
$\sigma_k ^{\max}$ for the polarisation angles, this drop can only be
explained by the removing of the $n\upi$-artefacts in the \pacman
map. That there is still a lot of noise in the map on small real space
scales (large $k$'s), which governs the power spectra on these scales,
can be seen from the dotted power spectrum, which was determined as
above while applying a simple window weighting in the
calculation. There is an additional difference of about one order of
magnitude between the two power spectra at large $k$-scales determined
for the two \pacman scenarios.

For comparison the power spectrum of pure white noise is
plotted as straight dashed line in Fig.~\ref{fig:psp2255e}. It can be
shown analytically, that white noise as observed through an arbitrary
window results in a spectrum of $\varepsilon_B(k) \propto k^2$.

As another independent statistical test, we applied the
\textit{gradient vector alignment statistic $\tilde{V}$} to the RM and
$\varphi^0$ maps calculated by the \pacman and the standard fit
algorithm. The gradients of RM and $\varphi^0$ were calculated using
the scheme as described in footnote 5 in
\citet{2003ApJ...597..870E}. For the two RM maps shown in
Fig.~\ref{fig:rmmaps}, we find a ratio of $\tilde{V}_{\rm
{stdfit}}/\tilde{V}_{\rm {pacman}} = 20$, which indicates a
significant improvement mostly resulting from removing the
$n\upi$-artefacts by the \pacman algorithm. The calculation of the
normalised quantity $V$ yielded $V_{\rm {stdfit}} = -0.75$ and $V_{\rm
{pacman}}=-0.87$. Smoothing the \pacman RM map slightly leads to a
drastic decrease of the quantities $\tilde{V}$ and $V$. This indicates
that the statistic is still governed by small scale noise. This can be
understood by looking at the RM maps of Abell 2255 in
Fig.~\ref{fig:rmmaps}. The extreme RM values are situated at the
margin of the source which is, however, also the noisiest region of
the source. Calculating the normalised quantity for the high
signal-to-noise region yields $V_{\rm {pacman}}=-0.56$. For
comparison, the normalised quantity for the standard fit of the same
high signal-to-noise region is $V_{\rm {stdfit}}=-0.94$.

\subsection{Hydra North \label{sec:hydranorth}}
The polarised radio source Hydra~A, which is in the centre of the
Abell cluster 780, also known as the Hydra~A cluster, is located at a
redshift of 0.0538 \citep{1991trcb.book.....D}. The source Hydra~A
shows an extended, two-sided radio lobe. Detailed X-ray studies have
been performed on this cluster \citep[e.g.][]{1997ApJ...481..660I,
1998MNRAS.298..416P, 2001ApJ...557..546D}. They revealed a strong
cooling core in the cluster centre. The Faraday rotation structure was
observed and analysed by \citet{1993ApJ...416..554T}. They reported RM
values ranging between \mbox{$-1\,000$ rad m$^{-2}$} and
\mbox{$+3\,300$ rad m$^{-2}$} in the north lobe and values down to
\mbox{$-12\,000$ rad m$^{-2}$} in the south lobe.
   
Polarisation angle maps and their error maps for frequencies at
$7\,815$, $8\,051$, $8\,165$, $8\,885$ and \mbox{$14\,915$ MHz}
resulting from observation with the Very Large Array were kindly
provided to us by Greg Taylor. A detailed description of the radio
data reduction can be found in \citet{1990ApJ...360...41T} and
\citet{1993ApJ...416..554T}.

\begin{figure}
\includegraphics[width=0.5\textwidth]{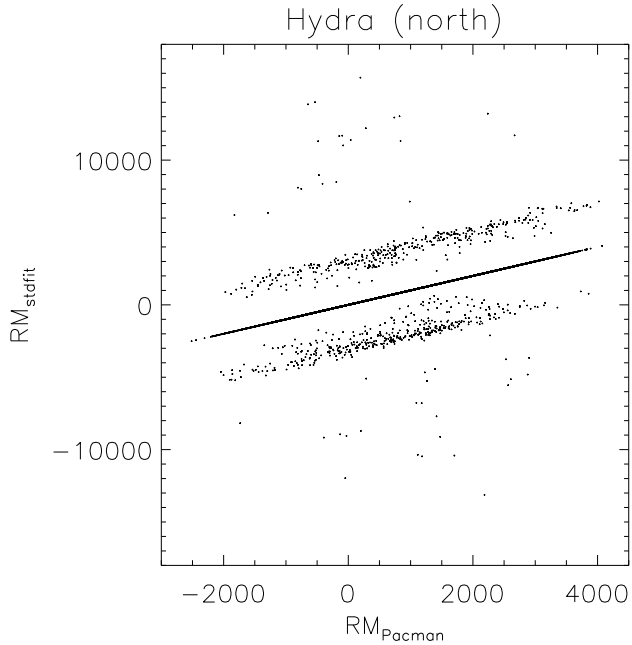}
\caption{A pixel-by-pixel comparison for the Hydra north lobe of
$RM_{\rm {stdfit}}$ values obtained using the standard fit plotted on
the $y$-axis versus the $\RM_{\rm {Pacman}}$ values calculated by
employing our new algorithm \textit{Pacman}. The parallel
scattered points at $\pm$ 3000-4000 rad m$^{-2}$ and $\pm\,10\,000$
rad m$^{-2}$ are a result of spurious solutions to the
$n\upi$-ambiguity obtained by the standard fit algorithm. For the
calculation of the RM maps, all five frequencies were used.}
\label{fig:comphydran}
\end{figure}

In this section, we concentrate on the north lobe of Hydra~A and
discuss the south lobe separately in Sect.~\ref{sec:hydrasouth}. The
south lobe is more depolarised, leading to a lower signal-to-noise
than in the north lobe. The north lobe is a good example of how to
treat noise in the data, while the south lobe gives a good opportunity
to discuss the limitations and the strength of our algorithm
\emph{Pacman}. 

Using the polarisation data for the four frequencies at around 8 GHz,
we calculated the standard fit RM map which is shown in the upper
middle panel of Fig.~\ref{fig:rmmaps}. The maximal allowed error in
polarisation angle was chosen to be \mbox{$\sigma_k ^{\max} =
25\degr$} and \mbox{RM$^{\max} = 15\,000$ rad m$^{-2}$}. A \pacman
RM map is shown in the middle of the middle panel in
Fig.~\ref{fig:rmmaps} below the standard fit map. This map was
calculated using the four frequencies at around 8 GHz with a
\mbox{$\sigma_{k=8GHz} ^{\max} = 30\degr$} and additionally where
possible the fifth frequency at 15 GHz was used with a
\mbox{$\sigma_{k=15GHz} ^{\max} = 35\degr$}. RM$^{\max}$ was chosen
to be \mbox{$15\,000$ rad m$^{-2}$}. From a visual comparison of the
two maps, one can conclude that the standard fit map looks more
structured on smaller scales and less smooth than the \pacman
map. This structure on small scales might be misinterpreted as small
scale structure of the RM producing magnetic field. Note that the
difference in the RM maps is mainly due to the usage of the fifth
frequency for the \pacman map demonstrating that all information
available should be used for the calculation of RM maps in order to
avoid misinterpretation of the data.

Since the four frequencies around 8 GHz are close together, a wrong
solution of the $n\upi$-ambiguity would manifest itself by differences
of about \mbox{$\Delta \RM = 10\,000$ rad m $^{-2}$}. Such jumps are
not observed in the main patches of the standard fit RM map shown in
Fig.~\ref{fig:rmmaps}. However, including the information contained in
the polarisation angle map of the 5th frequency at \mbox{15 GHz},
which is desirable as explained above, one introduces the possibility
of $n\upi$-ambiguities resulting in \mbox{$\Delta \RM =
3\,000...4\,000$ rad m$^{-2}$}. Therefore, a scatter plot of a
pixel-by-pixel comparison between a standard fit and a \pacman map,
calculated both using the additional available information on the
fifth frequency, is shown in Fig.~\ref{fig:comphydran}. The
parallel scattered points are spurious solutions found by the
standard fit algorithm. One can clearly see that they develop
at $\pm$ 3000-4000 rad m$^{-2}$ and less pronounced at $\pm\,10\,000$
rad m$^{-2}$. 

As we will discuss below, the 15 GHz data set may have a lower
signal--to--noise ratio than the 8 GHz data and thus, a lower error
threshold $\sigma_k ^{\max}$ will eliminate most of the wrong
fits. However, it is a particular strength of \pacman that it yields
still reliable results even if choosing larger error thresholds.

\begin{figure}
\includegraphics[width=0.5\textwidth]{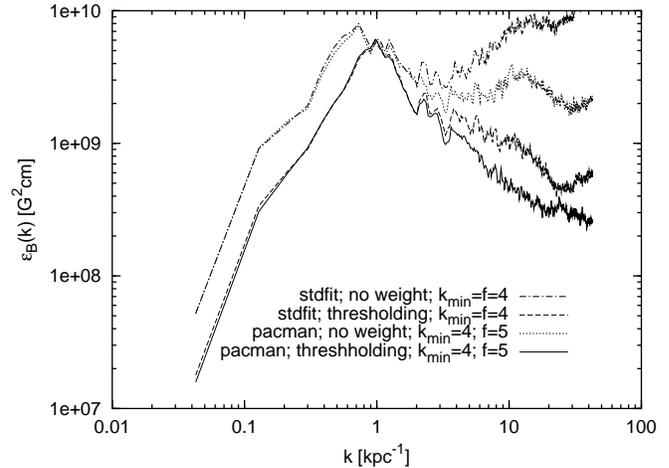}
\caption{Various power spectra calculated for the RM maps for the
north lobe of Hydra~A are shown. The dashed dotted line and the dotted
line represent the power spectra calculated from the standard fit RM
map and the \pacman RM map, respectively, while no window weighting
was applied. A threshold window weighting assuming a $\sigma_0 ^{\RM}$
= 75 rad m$^{-2}$ yield the dashed line power spectrum for the
standard fit map and the solid line for the \pacman map. The standard
fit maps were calculated from the four frequencies at 8 GHz whereas
the \pacman maps were determined using additionally the fifth
frequency when possible. For both algorithms, \mbox{$\sigma_k ^{\max}
= 35\degr$} was used. The influence of the small--scale pixel noise
on the power spectra can clearly be seen in this figure at large
$k$-values.}
\label{fig:ps_comp}
\end{figure}

In order to study the influence of the noise on small scales, we
calculated the power spectra from RM maps obtained using different
parameter sets for the \pacman and the standard fit algorithm. As for
Abell 2255E, all calculations of power spectra were done for RM areas
which would be covered by the \pacman fit if the same parameters were
used. Again this leads to exclusion of pixels in the standard fit RM
map which are not associated with the source. 

There is a clear depolarisation asymmetry of the two lobes of Hydra~A
observed as described by the Laing--Garrington effect
\citep{1993ApJ...416..554T, 2004AJ....127...48L}. Therefore, we assume
for the calculation of any power spectra for the Hydra source, that
the source plane is tilted by an inclination angle of 45$\degr$ where
the north lobe points towards the observer and the south lobe away
from the observer.

For a first comparison, we calculated the power spectra for two RM
maps similar to the one shown in Fig.~\ref{fig:rmmaps}. They were
obtained using the four frequencies at 8 GHz for the standard fit
algorithm and using additionally the fifth frequency when possible for
the \pacman algorithm. The other parameters were chosen to be
\mbox{$\sigma_k ^{\max} = 35\degr$} and $\RM^{\max}=15\,000$ rad
m$^{-2}$. For these two RM maps, we determined power spectra applying
firstly no window weighting at all and secondly threshold window
weighting as described in the end of Sec.~\ref{sec:powerspectra}. We
choose the threshold to be \mbox{$\sigma_0^{\RM}$ = 75 rad m$^{-2}$}
which represents the high signal-to-noise region. The respective
spectra are exhibited in Fig.~\ref{fig:ps_comp}.

For the power spectra, $n\upi$-artefacts should play only a minor role
in the calculation of the power spectra as explained above. Therefore,
any differences arising in these spectra are caused by the varying
treatment of noise in the map making process or in the
analysis. Comparing the power spectra without window weighting, the
one obtained from the \pacman RM map lies below the one from the
standard fit RM map. The difference is of the order of half a
magnitude for large $k$'s -- small real space scales. The standard fit
power spectrum seems to increase with larger $k$ but the \pacman one
seems to decrease over a greater range in $k$. The reason for this
difference is the same one which was responsible for the different
smoothness in the two RM maps shown in Fig.~\ref{fig:rmmaps}, namely
that only four frequencies were used for the standard fit RM map. Even
introducing a window weighting scheme, which down-weights noisy
region, cannot account entirely for that difference as can be seen
from the comparison of the window weighted power spectra. This is
especially true because the small-scale spatial structures are found
even in the high signal-to-noise regions when using only four
frequencies for the RM map calculation. This demonstrates again how
important it is to include all available information in the map making
process.

\begin{figure}
\includegraphics[width=0.5\textwidth]{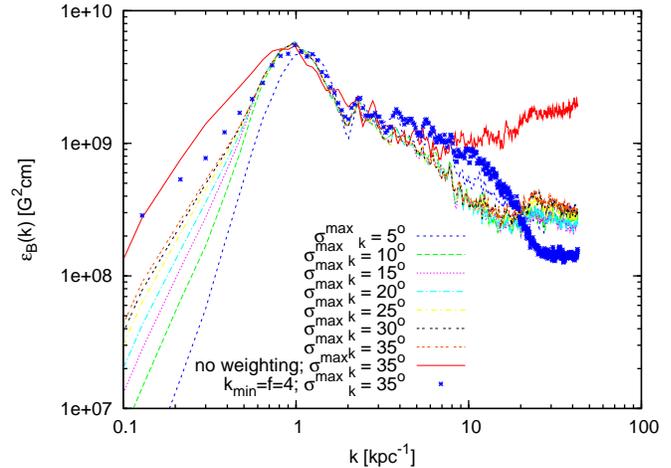}
\caption{Power spectra for the north lobe of Hydra~A calculated from a
series of \pacman RM maps are shown. The various coloured
dashed and dotted lines show power spectra from RM maps calculated for
$\sigma _k ^{\max}$ ranging from 5$\degr$ to 35$\degr$ using all five
frequencies $k_{\min}=5$. The spectra were determined applying
threshold window weighting ($\sigma_0 ^{\RM}$ = 50 rad m$^{-2}$). For
comparison the solid red line power spectrum, which is the
largest spectrum at $k>10$ kpc$^{-1}$, is not window weighted for
$\sigma _k ^{\max} = 35\degr$. The power spectrum represented by the
blue stars was derived from an \pacman RM map obtained using
the four frequencies at 8 GHz and a $\sigma _k ^{\max} = 35\degr$.}
\label{fig:ps_errprop}
\end{figure}

In order to study the influence of the maximal allowed measurement
error in the polarisation angle $\sigma _k ^{\max}$, we calculated the
power spectra for a series of \pacman RM maps obtained for $\sigma _k
^{\max}$ ranging from 5$\degr$ to 35$\degr$. The RM maps were
derived using all five frequencies $k_{\min}$= 5, not allowing any
points to be considered for which only four frequencies fulfil
\mbox{$\sigma_{k_{ij}} < \sigma_k ^{\max}$}. Again an
\mbox{$RM^{\max}=15\,000$ rad m$^{-2}$} was used. The resulting power
spectra using a threshold window weighting ($\sigma_0 ^{\RM}$ = 50 rad
m$^{-2}$) are exhibited in Fig.~\ref{fig:ps_errprop}. One can see that
they are stable and do not differ substantially from each other even
though the noise level increases slightly when increasing the $\sigma
_k ^{\max}$. This demonstrates the robustness of our \pacman
algorithm.

\begin{figure*}
\includegraphics[width=1.\textwidth]{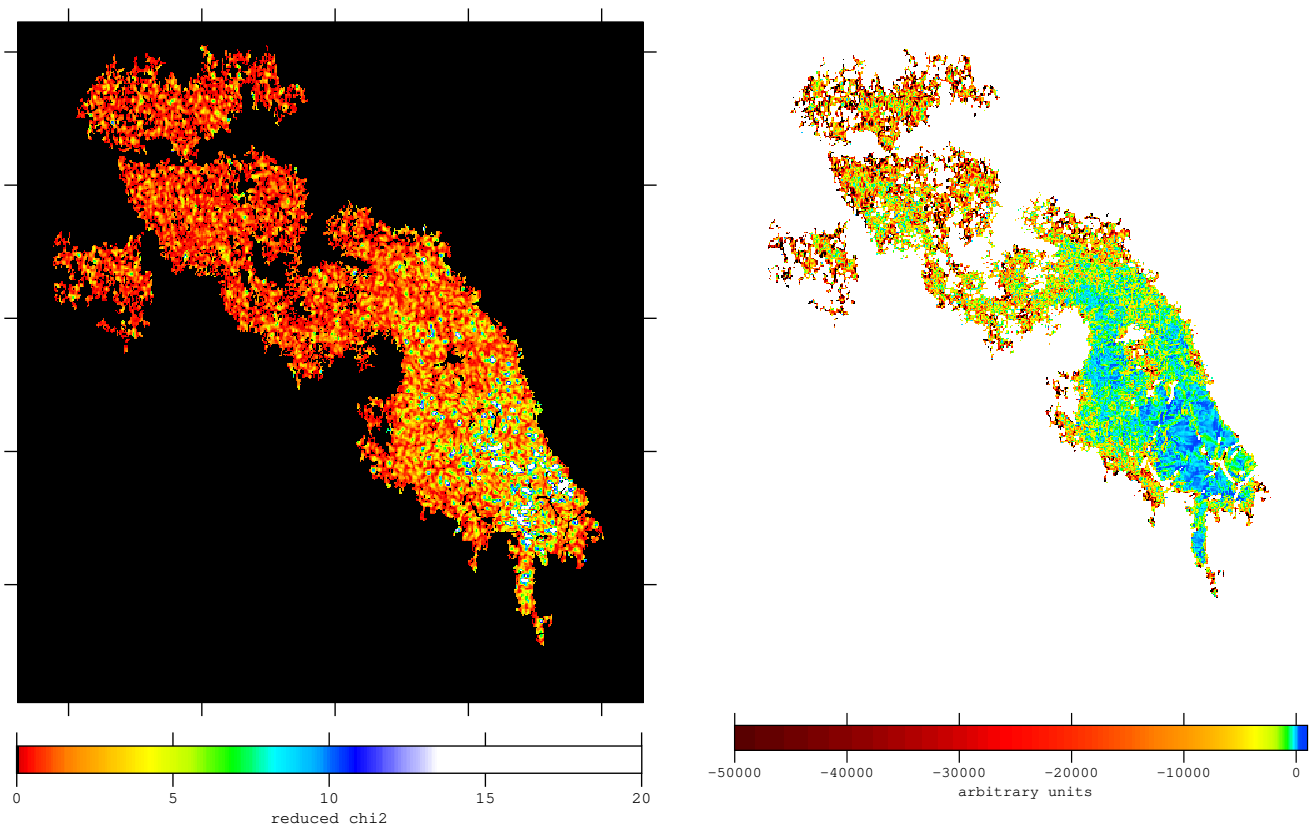}
\caption{On the left hand side a $\chi^2_{\nu{ij}}$ is
shown. The yellow and green regions have values larger than unity
indicating underestimation of errors. Note that for these regions the
RM fit used five frequencies whereas in the outer parts of the source
only four frequencies were used. A $\tilde{V}$ map for the north lobe
of Hydra~A is exhibited on the right panel. The extreme negative
values (red and black) correspond to strong anti-correlation of RM and
$\varphi^0$ fluctuations, which are produced by noise in the observed
maps. However, the green and blue coloured regions have values around
zero and are therefore high signal-to-noise regions (compare with the
$\sigma^{\RM}$ map in the middle panel of Fig.~\ref{fig:rmmaps}).}
\label{fig:Vhydra}
\end{figure*}

For comparison, we plotted two more power spectra in
Fig.~\ref{fig:ps_errprop}. For the solid red line spectrum,
no window weighting was applied to an RM map which was calculated as
above having \mbox{$\sigma _k ^{\max} = 35\degr$}. One can clearly see
that the spectrum is governed by noise on large $k$-scales -- small
real space scales. Thus, some form of window weighting in the
calculation of the power spectra seems to be necessary in order to
suppress a large amount of noise.

Another aspect arises in the noise treatment if we consider the power
spectra represented by the blue stars which was calculated using a
threshold window weighting ($\sigma_0= $ 50 rad m$^{-2}$). This power
spectrum represents the analysis for an RM map obtained using only the
four frequencies at 8 GHz while \mbox{$\sigma _k ^{\max} =
35\degr$}. It is striking that the noise level on large $k$-scales (on
small real space scales) is lower than for the RM maps obtained for the
five frequencies, one would expect it to be the other way
round. Apparently the fifth frequency has a lot of weight in the
determination of $\sigma^{\RM}_{ij}$ (see Eq.~(3) in paper I) since
this frequency is almost twice as large as the other four. If the
measurement errors of the polarisation angles are underestimated this
can lead to an underestimation of the uncertainty $\sigma^{\RM}_{ij}$
in the final RM value due to Gaussian error propagation. A more
accurate error estimate could be achieved by considering the error
treatment as described by \citet{1995MNRAS.273..877J}. They state in
their Appendix that errors are not uniform across synthesis images and
propose a different error algorithm.

One can test the hypothesis of underestimation of errors by
performing a $\chi^2_{\nu}$ test as described in
Sect.~\ref{sec:s_statistic}. The $\chi^2_{\nu_{ij}}$ map for the
\pacman RM map of Hydra North derived by using a multi-frequency fit
(see middle upper panel of Fig.~\ref{fig:rmmaps}) is exhibited in the
left panel of Fig.~\ref{fig:Vhydra}. The red values are close to
unity. However the yellow and blue regions in this map are values
larger than unity indicating that the errors are underestimated in
these regions. Note that in these regions there is also the fifth
frequency used for the RM fit which might indicate that the errors
for this frequency are underestimated by a larger factor than for the
other four frequencies.

We also applied the \textit{gradient vector product statistic
$\tilde{V}$} to the RM and $\varphi^0$ maps obtained by using the RM
maps as shown in the middle panel of Fig.~\ref{fig:rmmaps} and the
respective $\varphi^0$ maps. The calculation yielded for the ratio
$\tilde{V}_{\rm stdfit} /\tilde{V}_{\rm pacman}=1.2$. This corresponds
to a slight decrease of correlated noise in the \pacman maps. The
calculation of the normalised quantity resulted in $V_{\rm stdfit} =
-0.85$ for the standard fit and $V_{\rm pacman}=-0.83$ for the \pacman
maps. This is expected since the low signal-to-noise areas of the
north lobe of Hydra A will govern this gradient statistic and for both
RM maps these areas look very similar. In order to visualise this, we
calculated a $\tilde{V}$ map which is exhibited in
Fig.~\ref{fig:Vhydra}. The extreme negative values observed for the
low signal-to-noise regions (red regions) indicate strong
anti-correlation and thus, anti-correlated fluctuations of RM and
$\varphi^0$ in these regions. For the high signal-to-noise regions
(blue and green) moderate values varying around zero are
observed. Note the striking morphological similarity of the RM error
($\sigma^{\RM}$) map in the lower middle panel of
Fig.~\ref{fig:rmmaps} and of the $\tilde{V}$ map as shown in
Fig.~\ref{fig:Vhydra}. These two independent approaches to measure the
RM maps accuracy give a basically identical picture. These approaches
are complementary since they use independent information. The RM error
map is calculated solely from $\sigma_k$ maps, which are based on the
absolute polarisation errors, whereas the $\tilde{V}$ map is based
solely on the RM derived from the polarisation angle $\varphi(k)$
maps.

However, we note that the results for magnetic field strengths and
correlation lengths presented by \citet{2003A&A...412..373V} are not
changed substantially. This is owing to the fact that
\citet{2003A&A...412..373V} considered for the calculation of the
magnetic field properties only the power on scales which were larger
than the resolution element (beam). Therefore the small scale noise
was not considered for this calculation and thus, no change in the
results given is expected. The central magnetic field strength for
this cluster as derived during the course of this work is about 10
\mbox{$\umu$G} and the field correlation length is about \mbox{1 kpc}
being consistent with results presented in
\citet{2003A&A...412..373V}.

\subsection{The Quest for Hydra South\label{sec:hydrasouth}}
As already mentioned, the southern part of the Hydra source has to be
addressed separately, as it differs from the northern part in many
respects. One of them is that the depolarisation is higher in the
south lobe which especially complicates the analysis of this part. 

An RM map obtained employing the standard fit algorithm using the four
frequencies at 8 GHz is shown in the upper right panel of
Fig.~\ref{fig:rmmaps}. This RM map exhibits many jumps in the RM
distribution which seems to be split in lots of small patches having
similar RMs. Furthermore, RMs of \mbox{$-12\,000$ rad m$^{-2}$} are
detected (indicated as red regions in the map). If one includes the
fifth frequency at 15 GHz in the standard fit algorithm, the
appearance of the RM map does not change significantly, and most
importantly, the extreme values of about \mbox{$-12\,000$ rad
m$^{-2}$} do not vanish.

The application of the \pacman algorithm with conservative settings
for the parameters $\sigma_k ^{\max}$, $\sigma^{\Delta}_{\max}$ and
$g$, leads to a splitting of the RM distribution into many small,
spatially disconnected patches. Such a map does look like a standard
fit RM map and there are still the RM jumps and the extreme RM values
present. However, if one lowers the restrictions for the construction
of patches, the \pacman algorithm starts to connect patches to the
patch with the best quality data available in the south lobe using its
information on the global solution of the $n\upi$-ambiguity. An RM map
obtained pushing \pacman to such limits is exhibited in the middle
right panel of Fig.~\ref{fig:rmmaps}. Note that the best quality area
is covered by the bright blue regions and a zoom-in for this region is
also exhibited in the small box to the lower left of the source. In
the lowest right panel of Fig.~\ref{fig:rmmaps}, a $\sigma^{RM}$-map
is shown indicating the high quality regions by the red colour.

\begin{figure*}
\includegraphics[width=0.8\textwidth]{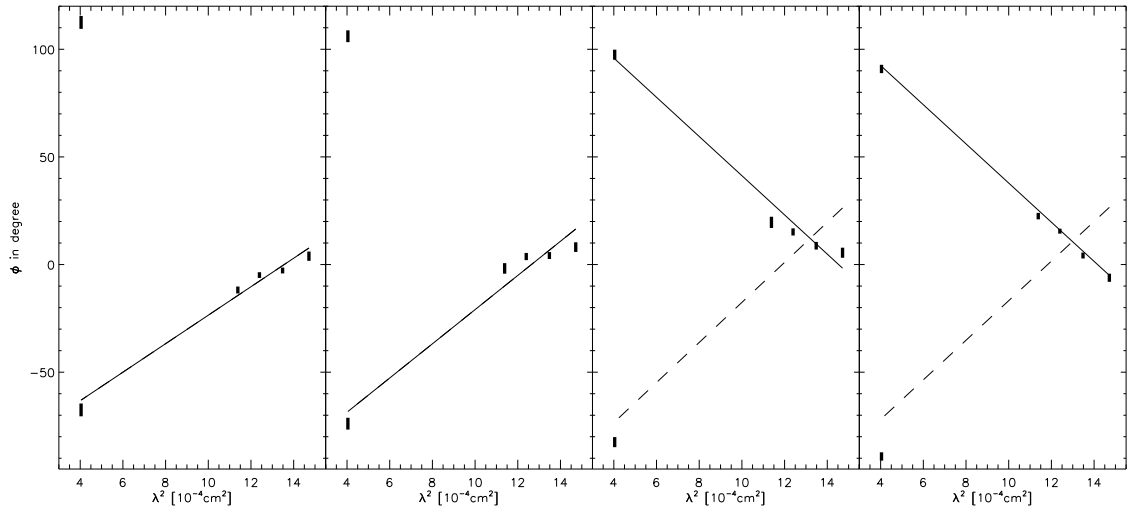}
\caption{An example for individual RM fits across an RM jump observed
in the RM maps of the south lobe of Hydra~A. The solid line indicates
the standard fit solution and the dashed one represents the \pacman
solution. The error bars in these plots indicate the polarisation
errors which were multiplied by~3. Note that this is an extreme
example, where \pacman chooses a solution to the individual fits in
noisy regions which has not necessarily the minimal $\chi^2$ value by
construction. However, the global statistical tests applied to the
whole RM map of Hydra South indicate that the \pacman map has less
artefacts in comparison to the standard fit RM map. We note, that for
this particular local example it is not trivial to decide which
algorithm is giving the right solution.}
\label{fig:rmplots}
\end{figure*}

The RM distribution in this \pacman map exhibits clearly fewer jumps
than the standard fit map although the jumps do not vanish
entirely. Another feature that almost vanishes in the \pacman map are
the extreme RMs of about \mbox{$-12\,000$ rad m$^{-2}$}. The RM
distribution of the \pacman map seems to be smoother than the one of
the standard fit map. However, if the individual RM fits for points,
which deviate in their RM values depending on the algorithm used, are
compared in a $\varphi(k)$-$\lambda_k^2$-diagram, the standard fit
seems to be the one which would have to be preferred since it does fit
better to the data at hand. \pacman has some resistance to pick this
smallest $\chi^2$ solution if it does not make sense in the context of
neighbouring pixel information. As an example, we plotted individual
RM fits which are observed across the RM jumps in
Fig.~\ref{fig:rmplots}.  However as shown above, one has to consider
that the measurement errors might be underestimated which gives much
tighter constraints on the fit than it would otherwise be.

\begin{figure}
\includegraphics[width=0.5\textwidth]{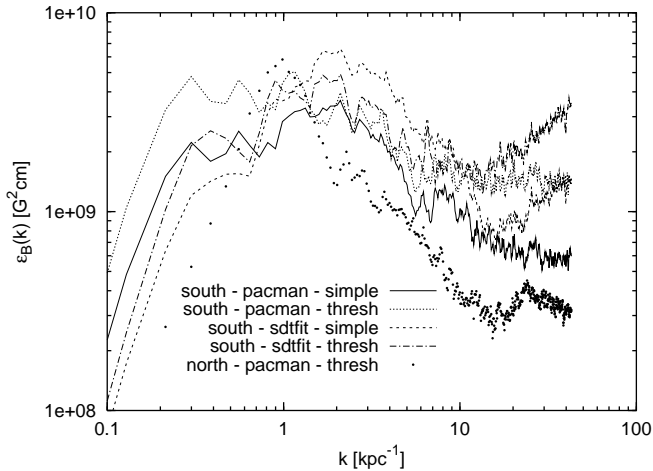}
\caption{Various power spectra for the RM maps of the south lobe of
the Hydra source are compared with the power spectrum of the north
lobe calculated from the \pacman RM map ($k_{\min}=5$, $\sigma_k
^{\max} = 35 \degr$) employing a threshold weighting with $\sigma_0
^{\RM}=50$ rad m$^{-2}$ represented as filled circles.}
\label{fig:pshydra_south}
\end{figure}

In order to investigate these maps further, we also calculated the
threshold weighted ($\sigma_0 ^{\RM}$ = 50 rad m$^{-2}$) and simple
weighted power spectra from the \pacman and the standard fit RM maps
of the south lobe and compared them to the one from the north
lobe. The RM maps used for the comparison were all calculated
employing $\sigma_k ^{\max}= 35\degr$ and $k_{\min}=f=5$. The
various power spectra are shown in Fig.~\ref{fig:pshydra_south}. One
can clearly see that the power spectra calculated for the south lobe
lie well above the power spectra from the north lobe, which is
represented as filled circles.

Concentrating on the comparison of the simple window weighted power
spectra of the south lobe in Fig.~\ref{fig:pshydra_south} (solid line
from \pacman RM map and dashed line from standard fit RM map) to the
one of the north lobe, one finds that the power spectrum calculated
from the \pacman RM map is closer to the one from the north lobe than
the power spectrum derived from the standard fit RM map. This
indicates that \pacman RM map might be the right solution to the RM
determination problem for this part of the source. However, this
difference vanishes if a different window weighting scheme is applied
to the calculation of the power spectrum. This can be seen by
comparing the threshold window weighted power spectra (dotted line for
the power spectra of the \pacman RM map and dashed dotted line for the
power spectra from the standard fit RM map) in
Fig.~\ref{fig:pshydra_south}. Since the choice of the window weighting
scheme seems to have also an influence on the result the situation is
still inconclusive.

An indication for the right solution of the $n\upi$-ambiguity problem
may be found in the application of our \textit{gradient vector product
statistic $\tilde{V}$} to the \pacman maps and the standard fit maps
of the south lobe (as shown in Fig.~\ref{fig:rmmaps}). The
calculations yielded a ratio $\tilde{V}_{\rm {stdfit}}/\tilde{V}_{\rm
{pacman}} = 12$. This is a difference of one order of magnitude and
represents a substantial decrease in correlated noise in the \pacman
maps. The calculation of the normalised quantity yielded $V_{\rm
{stdfit}} = -0.69$ for the standard fit maps and $V_{\rm {pacman}} =
-0.47$ for the \pacman maps. This result strongly indicates that the
\pacman map should be the preferred one.

The final answer to the question about the right solution of the RM
distribution problem for the south lobe of Hydra~A has to be postponed
until observations of even higher sensitivity, higher spatial
resolutions and preferably also at different frequencies are
available.

\section{Conclusions \& Lessons} \label{sec:conclusion}
We demonstrated the robustness of our \pacman algorithm which is
especially useful for the calculation of RM and $\varphi^0$ maps of
extended radio sources. It dramatically reduces $n\upi$-artefacts in
noisy regions and makes the unambiguous determination of RMs in these
regions possible. Any statistical analysis of the RM maps will profit
from this improvement.

In the course of this work the observational data taught us the
following lessons:

\begin{enumerate}
\item It is important to use all available information obtained by the
observation. It seems especially desirable to have a good frequency
coverage. The result is a smoother, less noisy map.

\item For the individual least-squares fits of the RM, it is advisable
to use an error weighting scheme in order to reduce the noise.

\item Global algorithms are preferable, if reliable RM values are
needed from low signal-to-noise regions at the edge of the source.

\item Sometimes it is necessary to test many values of the parameters
$\RM^{\max}$, $\sigma^{\max}_k$ and $\sigma^{\Delta}_{\max}$, which
govern the \pacman algorithm, in order to investigate the influence on
the resulting RM maps.

\item One has to keep in mind that whichever RM map seems to be most
believable, it might still contain artefacts. One should do a careful
analysis by looking on individual RM pixel fits but one also should
consider the global RM distribution. We presented the \textit{gradient
vector product statistic $\tilde{V}$} as a useful tool to estimate the
level of reduction of cross-correlated noise in the calculated RM and
$\varphi^0$ maps.

\item For the calculation of power spectra it is always preferable to
use a window weighting scheme which has to be carefully
selected. 

\item Finally, under or overestimation of the measurement errors of
polarisation angles (which will propagate through all calculations due
to error propagation) could influence the final results. We performed
a reduced $\chi^2_{\nu}$ test in order to test if the errors are over
or under estimated.
\end{enumerate}

We conclude that the calculation of RM maps is a very difficult task
requiring a critical view to the data and a careful noise
analysis. However, we are confident that \pacman offers a good
opportunity to calculate reliable RM maps and to understand the data
and its limitations better. Furthermore, such improved RM maps
and the according error analysis for Hydra North and Abell 2255 were
presented in this paper.

\section*{acknowledgements}
We want to thank Federica Govoni and Greg Taylor for allowing us to
use their data on Abell 2255 and Hydra to test the new algorithm. We
like to thank Tracy Clarke, James Anderson, Phil Kronberg and an
anonymous referee for useful comments on the
manuscript. K.~D.~acknowledges support by a Marie Curie Fellowship of
the European Community program "Human Potential" under contract number
MCFI-2001-01227.

\bibliographystyle{mn2e}
\bibliography{literature}

\label{lastpage}

\end{document}